\documentclass[aps,pra,twocolumn,showpacs,groupedaddress]{revtex4}  
\usepackage{graphicx}  
\usepackage{dcolumn}   
\usepackage{bm}        
\usepackage{amssymb}   
\usepackage{amsmath}
\usepackage{float}
\usepackage{ulem}
\usepackage{enumerate}
\begin{document}
	\title{Distinct collective states due to trade-off between attractive and repulsive couplings}
	\author{K. Sathiyadevi$^{1}$,  V. K. Chandrasekar$^{1}$,  D. V. Senthilkumar$^2$ and M. Lakshmanan$^{3}$}
	\address{$^1$Centre for Nonlinear Science \& Engineering, School of Electrical \& Electronics Engineering, SASTRA Deemed University, Thanjavur -613 401, Tamil Nadu, India. \\$^2$School of Physics, Indian Institute of Science Education and Research, Thiruvananthapuram-695016, India. \\ 
		$^3$Centre for Nonlinear Dynamics, School of Physics, Bharathidasan University, Tiruchirappalli - 620 024, Tamil Nadu, India. 
	}
	\begin{abstract} 
		\par    We investigate the effect of repulsive coupling together with an attractive coupling in a network of nonlocally coupled oscillators.  To understand the complex interaction between these two couplings we introduce a control parameter in the repulsive coupling which plays a crucial role in inducing distinct complex collective patterns. In particular, we show  the emergence of various cluster chimera death states through a dynamically distinct transition route, namely the oscillatory  cluster state and coherent oscillation death  state as a function of the  repulsive coupling in the presence of the attractive coupling. In the oscillatory  cluster state, the oscillators in the network are grouped into two distinct dynamical states of homogeneous and inhomogeneous  oscillatory states. Further, the network of coupled oscillators follow the same transition route in the entire  coupling range.  Depending upon distinct coupling ranges the system  displays different number of  clusters in the death state and oscillatory state.   We also observe that the number of coherent domains in the  oscillatory cluster state exponentially decreases with increase in coupling range and obeys a power law decay.  Additionally, we show analytical stability for observed solitary state, synchronized state and  incoherent oscillation death state.
		
	\end{abstract}
	\maketitle
	\section{Introduction}
	
	A network of  coupled nonlinear dynamical systems manifests itself into a  plethora of intriguing collective dynamical behaviors  such as clusters, pattern formation,  synchronization and so on \cite{int}.  Among them, oscillation queching is one of the intriguing phenomenon observed in various physical, chemical and biological systems \cite{od}. The phenomenon of  quenching  can be further distinguised as (i) amplitude death ($AD$)  and (ii) oscillation death ($OD$). 	$AD$ is the stabilization of already existing homogeneous steady state $(HSS)$ which was initially identified due to a large parameter mismatch \cite{od_r1}-\cite{od_r2}, but later observed in identical systems with meanfield diffusive interaction \cite{od_r5} and time delayed coupling \cite{od_r4}-\cite{add}.   In the $OD$ state, the oscillators in the network cease their oscillations under the coupling  and populate in at least two inhomogeneous steady states $(IHSS)$.    Oscillation death was  also initially observed due to parameter mismatch in coupled systems \cite{iprl1968}.  Later, the phenomenon of quenching of  oscillations  was shown to emerge even in identical oscillators with time-delayed interactions \cite{od_sen1}-\cite{od_sen2} and eventually realized in a variety of couplings, such as dynamic coupling \cite{od_rr4}, conjugate coupling \cite{od_rr5}, environmental coupling \cite{od_rr6} and in repulsive mean field interactions \cite{od_rr7}-\cite{ull}. $OD$ state has also been experimentally observed in chemical oscillators \cite{od_app1}, chemical droplets \cite{od_app2}, thermokinetic oscillators \cite{od_app3} and electronic circuits \cite{od_app4}.   Multicluster $OD$ state was reported recently in an ensemble of globally coupled Josephson junctions \cite{clust_od}. 
	
	\par Chimera state, which corresponds to the coexistence of coherent and incoherent domains of oscillations in an identical network,  is  another emerging phenomenon that is being widely reported  both theoretically and experimentally in the recent literature \cite{chi1}-\cite{chi4}. In the recent past much attention has been paid towards understanding the onset  of such chimera states  \cite{t2}-\cite{t5}. Initially such states were reported under nonlocal coupling \cite{chi_1}-\cite{gopal1}, eventually realized even in global coupling \cite{chi_4}-\cite{chen1}, and in nearest neighbour couplings \cite{chi_5}-\cite{chi_8}, as chimera states  were shown to share strong resemblance  to  (and can reveal underlying dynamical mechanisms in) many real world phenomena such as  unihemispheric sleep (i.e. ability of some mammals and birds sleeping with one half of the brain while  the other half remains awake) \cite{a1},  epileptic seizure \cite{a2}, neuronal bump states {\cite{a3},\cite{a4}}, power grids \cite{a5} and social systems \cite{a6}. Despite the existence of  substantial reports on the $OD$ and  the chimera states, both these states were reported as  separate dynamical entities until recently \cite{cd1}-\cite{prem2}.
	
	In this article, we will unravel the emergence of the fascinating phenomenon of chimera death state, which is characterized by the combined features of chimera and oscillation death state.   In the chimera death state, the oscillators in the network segregate into two coexisting domains, wherein one domain neighboring nodes occupy the same branch of the inhomogeneous steady state (spatially coherent $OD$) while in the other domain neighboring nodes are randomly distributed among the different branches of  the inhomogeneous steady state (spatially incoherent $OD$). The   inter-connection between the chimera and  the oscillation death states was reported by Zakharova {\it et al. } \cite{cd1}  using  a symmetry breaking nonlocal coupling, where the  transition from amplitude chimera to chimera death via in-phase synchronized state  was reported \cite{cd1}-\cite{cd3}.   Recently,  distinct types of chimera death states were also reported by Premalatha {\it et al. } \cite{prem1}.   It was  shown that the presence of nonisochronicity parameter leads to  structural changes in the chimera death region giving rise to the existence of different types of chimera death states such as multi-chimera death state, type-I periodic chimera death $(PCD-I)$ state and type-II periodic chimera death $(PCD-II)$ state \cite{prem1},\cite{prem2}.  
\par In this manuscript,   we consider a   network of nonlocally coupled van der Pol (vdP) oscillators with combined attractive and repulsive interactions.  It is known that  the counteracting  effects of  attractive-repulsive couplings  play a predominant role in  various chemical and biological processes.   For instance, excitation-contraction (EC) coupling was employed in cardiac myocytes \cite{qu} and a pair of neurons in the presence of coexisting excitatory (attractive) and inhibitory (repulsive) synaptic couplings was analyzed by  T. Yanagita {\textit{et. al} \cite{neu}}.  Further, in the gene regulatory network, positive and negative feedback loops are  often used to perform various functions such as  bistable switches, oscillators, and excitable devices \cite{gene}.  Here, we elucidate the emergence of various complex collective patterns due to the combined presence of attractive and repulsive couplings. We begin our analysis with a minimal network of two coupled vdP oscillators and  illustrate the onset of oscillation death as a function of the repulsive interaction. Further, we extend our analysis to a network  of coupled vdP oscillators with nonlocal  attractive-repulsive couplings and demonstrate the emergence of distinct collective dynamics as a function of the strength of the repulsive interaction. In particular,  the existence of chimera death preceded by a distinct collective state, namely oscillatory cluster state $(OC)$, will be demonstrated.  The oscillatory cluster is characterized by the coexisting homogeneous and inhomogeneous oscillatory states.   Finally we will establish that the  chimera death state occurs via the distinct oscillatory  cluster  state   due to  the interplay of  the nonlocal repulsive and attractive couplings using two parameter phase diagrams.
	 \par The structure of the paper is organized as follows:  In Sec. II, we describe our model of a network of nonlocally coupled van der Pol oscillators with combined  attractive and repulsive couplings. In Sec. III, we demonstrate  the emergence of oscillation death as a function of the repulsive  coupling in two coupled van der Pol oscillators. Further, we investigate the emergence of distinct chimera death state via oscillatory cluster state in a network of oscillators in Sec. IV and we discuss the global dynamical  behavior in Sec V. Finally, we summarize our results in sec. VI.

	\section{ The model}
	We consider a simple,  prototype, self-excitatory model of van der Pol (vdP) oscillators which can be constructed experimentally using electronic circuits that mimics the dynamics of the human heart \cite{vdp1}-\cite{vdp2}. Now, we consider a  network of nonlocally coupled  van der Pol oscillators with combined  attractive and repulsive interactions, whose governing equations are represented as
	\begin{eqnarray}
	\dot{ x}_{i} &=& {y}_{i} + \frac{\epsilon}{2P}\sum_{j=i-P}^{i+P}({ x}_{j}-{ x}_{i}), \nonumber \\ 
	\dot{ y}_{i} &=&\alpha (1-{ x}^2_{i}) { y}_{i} -{ x}_{i} -\frac{q\epsilon}{2P}\sum_{j=i-P}^{i+P}({y}_{j}-{ y}_{i}),
	\label{model}
	\end{eqnarray}
	where, $i = 1,2,...,N$. Here, $N$ is the total number of oscillators in the network.  In (\ref{model}), $\alpha$ is the damping parameter which manifests itself nearly sinusoidal oscillations for  smaller values and relaxation oscillations for  larger values. The nonlocal interaction  is controlled through the coupling strength $\epsilon$ and the coupling range (coupling radius) $r$ which is defined as $P$/$N$, where $P$ corresponds to the  total number of neighbours in both  the directions of  oscillators in the network.  Particularly, 
	the repulsive interaction among the oscillators is controlled through the parameter $q$. Initial conditions  for $x$ and $y$ are uniformly distributed   between $-1.0$ to $+1.0$. Runge-Kutta fourth order integration scheme is used with a time step of  $0.01$ for all our simulations. 
	
 Over the decades several investigations employing van der Pol oscillators have reported distinct dynamical behaviors under a variety of
coupling configurations.  
In particular, among the variety of collective behaviors  reported so far in  the literature using the coupled  vdP oscillators,  the phenomena of oscillation death, cluster
formation, chimera and chimera death will be reported in a single framework in the present manuscript. Further, we will also show 
the existence of chimera death preceded by a distinct collective state, namely oscillatory cluster state $(OC)$
which is reported here for the first time in the literature to the best of our knowledge.  Swing of synchronized state
is also observed without nonisochronicity parameter in contrast to the one reported in the literature. Further, systematic
bifurcation analysis of different dynamical transitions has also carried out all through the article.

\section{Emergence of oscillation death in two coupled Van der Pol oscillators}
	
	\begin{figure*}
		\centering
		\includegraphics[width=19.0cm]{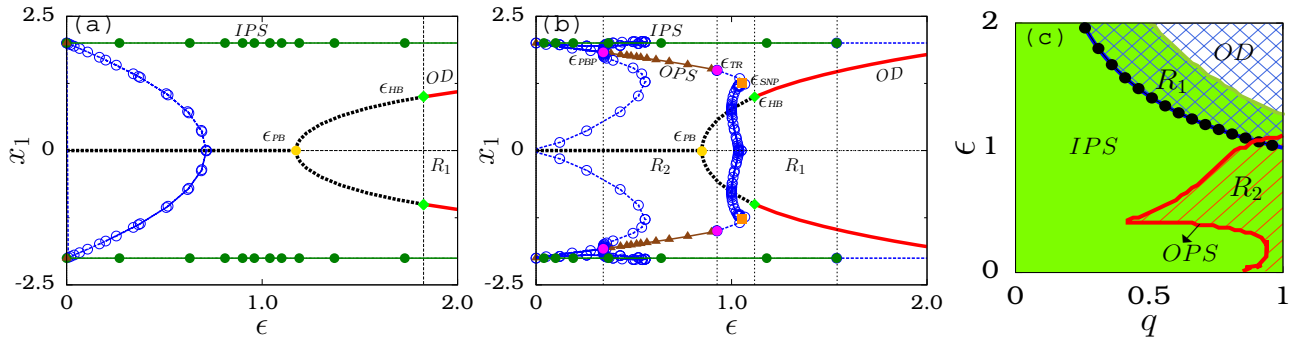}
		\caption{(color online) (a)-(b) show bifurcation diagrams  (using XPPAUT) for $N=2, \alpha=0.5$.   Fig. (a) is plotted for  $q= 0.3$  and (b)  for $q=0.8$. Filled circles (green) and triangles (brown) represent  stable $IPS$  and $OPS$  states, respectively. The unfilled circles correspond to unstable $OPS$. The  dotted (black) line and solid (red)  line depict unstable and stable nature of steady states. $TR$,  $PB$ and   $HB$ denote torus, pitchfork, and Hopf bifurcation points   respectively.  $PBP$ and $SNP$ denote the pitch fork and saddle node bifurcations of periodic orbits.  (c) Phase diagram in  $(q,\epsilon)$ space. $R_1$ and $R_2$ are the multistability regions of $IPS-OD$ and $IPS-OPS$, respectively.}
		\label{n2}
	\end{figure*}
	
	At first, we consider the limiting case of two identical van der Pol oscillators with  attractive and repulsive couplings between them.  The dynamical  transitions of the two coupled vdP oscillator will be analysed through the bifurcation diagrams {(using XPPAUT)} depicted in Figs.~\ref{n2}(a)-(b). The coupled  system (\ref{model}) is characterized by
	the following fixed points: (i) trivial fixed point: (0,0,0,0) and (ii) nontrivial fixed points :  $(x_1^*,y_1^*,-x_1^*,-y_1^*)$, where, $x_1^*=-\sqrt \frac{-1+\alpha \epsilon + q \epsilon^2}{\alpha\epsilon}$ and $y_1^*=\epsilon x^*$.  The corresponding eigen values are
	\begin{eqnarray}
	\lambda_{1,2} &=& \frac{1-\epsilon^2 \mp \sqrt{1+\epsilon^2 d_1-8 \epsilon^3 \tilde\alpha}}{2 \epsilon}, \nonumber\\
	\lambda_{3,4} &=& \frac{1-q\epsilon^2 \mp \sqrt{1+\epsilon^2 d_2-8 \epsilon^3 \tilde \alpha}}{2\epsilon},
	\label{eig}
	\end{eqnarray}
	where  $\tilde{\alpha}=\alpha+q\epsilon$, $d_1 = 6+\epsilon^2$ and $ d_2 = 4-2q+q^2\epsilon^2$.
	From an analysis of the above eigen values, we find that a  pitchfork bifurcation (PB) occurs at the critical coupling strength $\epsilon_{PB} = \frac{-\alpha+\sqrt{\alpha^2+4q}}{2q}$. 
	The unstable inhomogeneous steady state $(IHSS)$ which emerges through a  symmetry breaking pitchfork bifurcation $(PB)$ is  stabilized via the  Hopf bifurcation point $\epsilon_{HB} = \frac{1}{\sqrt q}$  which can be deduced by equating the real parts of the eigen values to zero.	Figure~\ref{n2}(a) is depicted for the repulsive coupling $q=0.3$ as a function of the nonlocal coupling strength $\epsilon$. For feeble values of the repulsive coupling $(q)$,  in-phase synchronized $(IPS)$ state (represented by lines connected by filled circles) is found to be stable in the explored range of $\epsilon$ due to the predominant effect of the attractive coupling over the  repulsive coupling. Dotted line connected by unfilled circles correspond to unstable out-of-phase synchronized state $(OPS)$. By increasing  $\epsilon$ from zero, an unstable inhomogeneous  steady state (indicated by broken lines) emerges via   pitchfork  bifurcation  at  $\epsilon_{PB}=1.18 $ and  is stabilized through a Hopf bifurcation at  $\epsilon_{HB}=1.82$ thereby rendering the stable $IHSS$ (denoted by solid red lines)  to coexist with stable $IPS$ state in the region $R_1$ in Fig. \ref{n2}(a).
	
	\par Now, we increase the strength of the repulsive coupling to $q=0.8$. The dynamical transitions as a function of the nonlocal coupling  for $q=0.8$   are  depicted in the bifurcation diagram in Fig. \ref{n2}(b). It is evident that for smaller values of $\epsilon$,  the stable $IPS$ oscillations (lines connected by filled circles)  coexist with unstable out-of-phase synchronized $(OPS)$  oscillations (lines connected by unfilled circles).  Upon increasing  $\epsilon$ further, the trade-off between  the attractive and repulsive couplings result in the stabilization of the unstable $OPS$  via a pitchfork bifurcation of periodic orbit ($PBP$) at $\epsilon_{PBP}=0.344$   leading to bistability between  the stable $IPS$ and stable $OPS$ (indicated by lines connected by filled triangles) states in the region $R_2$.  The phase difference between the $OPS$ state is found to be $\pi$ and hence it may also be regarded as
antiphase synchronization.  By increasing the coupling strength further results in destabilization of stable $OPS$ states via   torus $(TR)$ bifurcation  at $\epsilon_{TR}=0.926$.  Further, the saddle-node bifurcation ($SNP$) of periodic orbits   occurs at $ \epsilon_{SNP}=1.04$, shown by a pair of squares in Fig. 1(b),  where the unstable quasiperiodic and 
 periodic  oscillations collide and disappear leading to  monostable limit cycle oscillation ($IPS$ state) in a narrow range of $\epsilon\in(0.926,1.12)$.
 In addition, the unstable inhomogeneous steady state  that emerges at $\epsilon_{PB}=0.845$ is stabilized via the Hopf bifurcation at $\epsilon_{HB}=1.12$ resulting in bistability between $IPS$ and $OD$ state in the region $R_1$.   The coupled system settles at the stable $OD$ state for strong nonlocal coupling as  evident from Fig. \ref{n2}(b). 
	
	\par  To observe the role of the repulsive coupling in inducing the oscillation death  in the simplest network of two coupled vdP oscillators, we have plotted the two  phase diagram in $(q,\epsilon)$ space in Fig. \ref{n2}(c). It elucidates that the coupled system exhibits only $IPS$ state in the entire range of $\epsilon$ for lower  values of the repulsive coupling strength $q$. At strong coupling limits the competition  among the attractive and the repulsive interactions leads to stable  $OD$ state.  Moreover, we find bistability between $IPS-OPS$ and  $IPS-OD$ in the $R_1$ and $R_2$ regions, respectively, in Fig. \ref{n2}(c). From the above analysis we find that a strong competition between the attractive and  repulsive interactions can give rise to the onset of oscillation death. It is also to be noted that the $OD$ state emerges only for larger repulsive coupling strengths.  Linear stability analysis  shows that the  $OD$ state is stable in the region $\epsilon_{HB} >\frac{1}{ \sqrt{q}}$ for $ q \le 1$	where the stabilization occurs through the Hopf bifurcation.  The analytical critical stability curve  across which a switch in the stability of unstable inhomogeneous steady state occurs as function of $q$   is represented by the line connected by filled circles in Fig. \ref{n2}(c).  In order to study the role of the repulsive coupling in inducing various other collective dynamics and the transition to chimera death state via the oscillatory cluster state we extend our analysis to a network of vdP oscillators with nonlocal attractive and repulsive couplings.

	\section{Role of repulsive interaction in a larger network of oscillators}
	In this section we study the effect of nonlocal repulsive  coupling together with an attractive coupling in a network of oscillators with $N=100$ for  the nonlocal coupling radius $r=0.3$.   
	\subsection{Swing of synchronized states}
	\begin{figure}[htb!] 
		\centering
		\hspace{-0.1cm}
		\includegraphics[width=8.00cm]{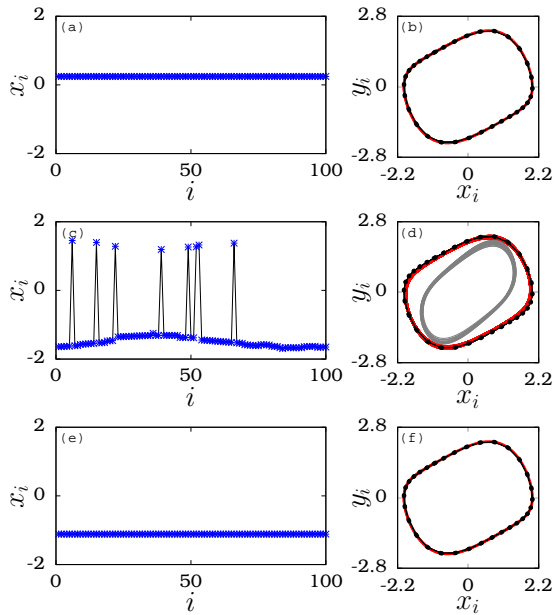}
		\caption{(color online) Snapshots for the variables $x_i$  and phase portraits of the system at $\alpha=0.5$,  $q=0.5$.  (a)-(b) synchronization ($SYN$) for $\epsilon=0.1$, (c)-(d) solitary state ($SS$)  for  $\epsilon=0.26$ and (e)-(d) synchronization ($SYN$) for $\epsilon=0.6$. The filled circles (black) connected by continuous line represent phase  trajectory of the uncoupled oscillator.}
		\label{swing}
	\end{figure} 
	\begin{figure}[htb!] 
		\centering
		\hspace{-0.1cm}
		\includegraphics[width=9.0cm]{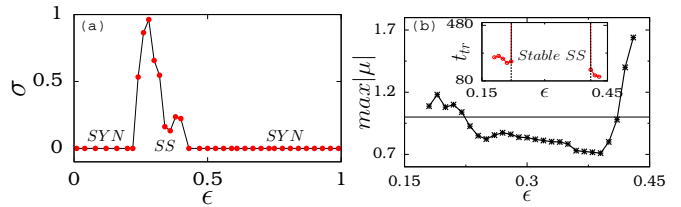}
		\caption{(color online) (a) Standard deviation ($\sigma$) at $\alpha=0.5$,  $q=0.5$ by varying coupling strength $\epsilon$,  the null value of the standard deviation  stands for synchronized state and nonzero values correspond to the solitary state and  (b) Maximum Floquet exponent for solitary state. max$|\mu|<1$ represents stable solution of periodic  orbits and max$|\mu|>1$ indicates the unstable nature of the corresponding dynamical state. The corresponding transient behavior is shown in the inset.  }
		\label{swing1}
	\end{figure} 
	
	For  the  repulsive coupling  strength $q = 0.5$,  we have found a swing like behavior of synchronized states as a function of $\epsilon$ (see Fig. \ref{swing}). The coupled system exhibits stable  synchronized oscillations for smaller values of $\epsilon$, the corresponding snapshot and phase portrait of which are shown in Figs. \ref{swing}(a)-(b) for $\epsilon=0.1$.  The solid (red)  line represents the phase portrait of globally synchronized oscillators whereas the filled circle points (black) connecting by the continuous line denote the phase portrait of uncoupled vdP oscillators and it is evident from the figures that the globally synchronized oscillators follow the original trajectory  of the uncoupled vdP oscillators. 
	
	By increasing $\epsilon$ further,  stable solitary state emerges as a result of destabilization of the globally synchronized state, as illustrated in Figs. \ref{swing}(c)-(d) for $\epsilon=0.26$.   From the figure, it is clear that the symmetry of the coupled system is broken spontaneously upon increasing the coupling strength  resulting in two different groups comprising of coherent and solitary oscillators.  The coherent oscillators (indicated by continuous red line) oscillate about the origin like the uncoupled oscillators whereas the solitary oscillators (indicated by solid grey line) oscillate with different amplitudes. Further increase in $\epsilon$ leads to a stabilization of completely synchronized state (see  Fig. \ref{swing}(e)) for  $\epsilon=0.6$ and it  follows the  same trajectory as that of the uncoupled oscillators.  Thus, we have observed  a swing of the synchronized state which was destabilized by the emergence of solitary state and again stabilized as a function of the coupling strength $\epsilon$.  It is to be noted that the coherent oscillators always evolve with the same amplitude and frequency as that of the uncoupled oscillator. In order to characterize the observed dynamical transition, we have estimated the standard deviation as used by Daido and Nakanishi \cite{daido} and 
	Premalatha {\it et. al.} \cite{prem2} using the formula,
	\begin{eqnarray}
	\sigma =  \langle (\overline{|x_i-\overline{x_i}|^2})^{1/2} \rangle_t,
	\label{lst}
	\end{eqnarray}
	where the bar represents the  ensemble average   and  $<\cdot>_t$ represents the time average. The standard deviation is depicted in  Fig. \ref{swing1}(a) as a function of 
	the nonlocal attractive coupling $\epsilon$. From Fig. \ref{swing1}(a),   it is evident that $\sigma$ takes null value for the synchronized state and nonzero value 
	for  the solitary state thereby corroborating the re-emergence of the stable synchronized state after the emergence of solitary state in a short range of $\epsilon$.  It is to be noted that nonisochronicity induced swing of synchronized state was reported in \cite{daido},\cite{prem2} whereas, in contrast,   we have observed the swing of the synchronized state as a function of the nonlocal coupling strength in the presence of the repulsive coupling without introducing any nonisochronicity parameter. As the oscillations in the solitary states are periodic and they are of same frequency, we can find the stability of these states using Floquet theory.  For this purpose, we have perturbed Eq. (\ref{model}) in the form $x_i=x_i^*+\eta_i$ and $y_i=y_i^*+\zeta_i$ and 
	derived the equations
	\begin{eqnarray}
	\small 	\dot{ \eta}_{i} &=& { \zeta}_{i} + \frac{\epsilon}{2P}\sum_{j=i-P}^{i+P}({\eta}_{j}-{ \eta}_{i}),  \nonumber\\ 
	\dot{ \zeta}_{i} &=&\alpha (1-{ x_i^*}^2) { \zeta}_{i} -(1+2x_i^*y_i^*){ \eta}_{i} -\frac{q
		\epsilon}{2P}\sum_{j=i-P}^{i+P}({ \zeta}_{j}-{\zeta}_{i}), \nonumber\\ 
	\label{floq}
	\end{eqnarray}
 where $x_i^*$ and $y_i^*$ correspond to the solitary state and  $\eta_i$ and $\zeta_i $ are the perturbation terms. The stability of the periodic orbits can be determined from the values of the Floquet multipliers ($\mu_i$, i=1,2,...N).  Whenever $\eta_i$ and $\zeta_i$ asymptotically approach zero or finite values, all the Floquet multipliers $\mu_i$'s should lie within a unit circle in the complex plane (or $|\mu_i| < 1$). Usually one of the values of $\mu_i$ can take the value $1$, which is referred to as Goldstone mode in the literature. In such situations, the periodic orbit is stable.  If any one of the $\mu_i$'s is greater than one,   the functions $\eta_i$ and $\zeta_i$ are found to grow up and consequently the periodic orbits are not stable.  We have plotted the maximum value of Floquet exponents $(max|\mu_i|)$ as a function of the coupling   strength $\epsilon$ in Fig. \ref{swing1}(b). It is clear that the value of  $max|\mu| < 1$ shows stable region of solitary state and  $max|\mu| > 1$ indicates the unstable region.    The corresponding transient behavior is shown in  the inset of Fig. \ref{swing1}(b) which clearly depicts that the values of the  Floquet exponents increase with decreasing transient time. In the stable solitary state the system experiences a very large transient  time than the unstable region  where the transient solitary state exists only for a  finite time. 
	
	\subsection{Quasi-periodic chimera}
\begin{figure}[htb!] 
		\centering
		\hspace{0.50cm}
		\includegraphics[width=8.5cm]{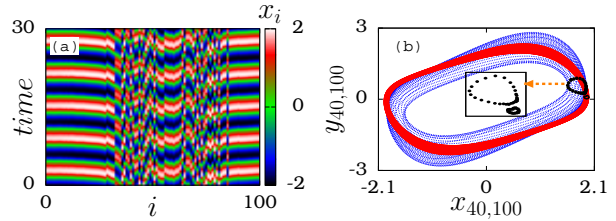}
		\caption{(color online) (a) Space-time plot of the chimera state at $\alpha=0.5$, $\epsilon=0.3$ and $q=0.77$. (b) shows  the phase portrait of  representative oscillators from the  coherent   and incoherent groups. The red trajectory corresponds to the representative oscillator  ($x_{100},y_{100}$) from the coherent group and that from the incoherent group ($x_{40},y_{40}$) is  represented by blue trajectory. The corresponding Poincar\'e points are shown on the phase portrait trajectory with filled circles and star points.  The closed loop of the  Poincar\'e points confirms the quasi-periodic nature of chimera state }
		\label{chi}
	\end{figure}

	A slight increase in $q$, breaks the system symmetry spontaneously thereby rendering  one group of  oscillators to the coherent region and the other group of oscillators to
	the incoherent state leading to the genesis of the fascinating hybrid  state of chimera ($CM$). The space-time plot and phase portraits  of quasi-periodic chimera state for $\alpha=0.5$, $\epsilon=0.3$ and $q=0.77$ are illustrated in Figs. \ref{chi}(a)-(b). The oscillators are found to exhibit quasi periodic oscillations in both  coherent and incoherent regions. Representative oscillators from both the groups are shown in Fig. \ref{chi}(b).  Dotted (blue) and  solid  (red) lines correspond to  the oscillator  $x_{100}$ from the coherent group and $x_{40}$  from the incoherent group, respectively. The Poincar\'e points (filled circles and star points) are depicted on the  phase  trajectory to corroborate that their closed loop structure which signify the quasi-periodic nature of oscillations in the chimera state.  The enlarged image of the Poincar\'e trajectory is shown in   Fig. \ref{chi}(b).
	\subsection{Collective dynamics at maximum repulsive coupling $(q=1)$}
	\begin{figure}[htb!] 
		\centering
		\vspace{0.5cm}
		\includegraphics[width=8.50cm]{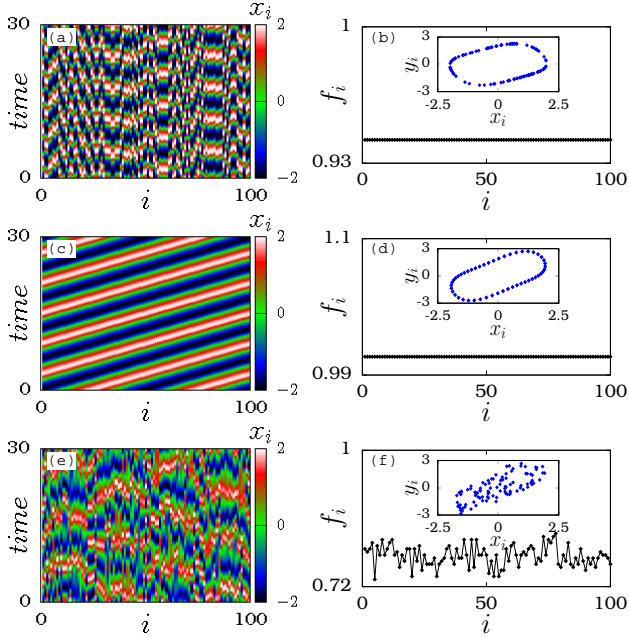}
		\caption{(color online)  Space-time plots  and  the corresponding frequencies  $f_i$ for   (a)-(b) desynchronization-$I$ ($DS-I$) at $\epsilon=0.05$, (c)-(d) travelling wave ($TW$) at $\epsilon=0.5$  and (e)-(f) desynchronization-$II$ ($DS-II$) at $\epsilon=0.92$. Other parameters are $\alpha=0.5$ and $q=1$.}
		\label{des}
	\end{figure}
	In this section, we study the dynamics at  the maximum repulsive coupling strength ($q=1$).   Space-time plots and the corresponding frequencies $f_i$ are depicted in Fig. \ref{des} for $\alpha=0.5$ as a function of the nonlocal coupling strength $\epsilon$. The coupled oscillator network evolves independently resulting in desynchronization-I  ($DS-I$) state for $\epsilon=0.05$ (see Figs. \ref{des}(a)-(b)). In this case,  the competition between the  attractive and  the repulsive interactions leads to the desynchronized state with same amplitude and frequency but with different phases (See Fig. \ref{des}(a)) . The phases of the oscillators (even though they oscillate with same frequency) are distributed  randomly between zero and $2\pi$. The inset in Fig. \ref{des} (b) shows the snapshot image of $DS-I$ state.  By increasing  $\epsilon$, we observe that the oscillators in the desynchronized group align spontaneously with constant velocity and form a stable coherent travelling wave ($TW$)  as depicted in Figs. \ref{des}(c)-(d) for $\epsilon=0.5$. In contrast to  the desynchronized state,  we find that the oscillators evolve  with constant phase differences distributed  between zero and $2\pi$ as depicted in the inset of Fig. \ref{des}(d).  Further increase in $\epsilon$ leads to strong competition between  the attractive and the repulsive nonlocal couplings resulting in desynchronized state ($DS-II$) as  shown in Figs. \ref{des} (e)-(f) for $\epsilon=0.92$. The amplitude, phase and frequency of all the oscillators are found to evolve independently as is evident from Figs. \ref{des}(e)-(f).   The  inset in Fig. \ref{des}(f) shows the snapshot of completely desynchronized state.  
	
	\begin{figure}[htb!] 
		\centering
		\hspace{-0.1cm}
		\includegraphics[width=9.00cm]{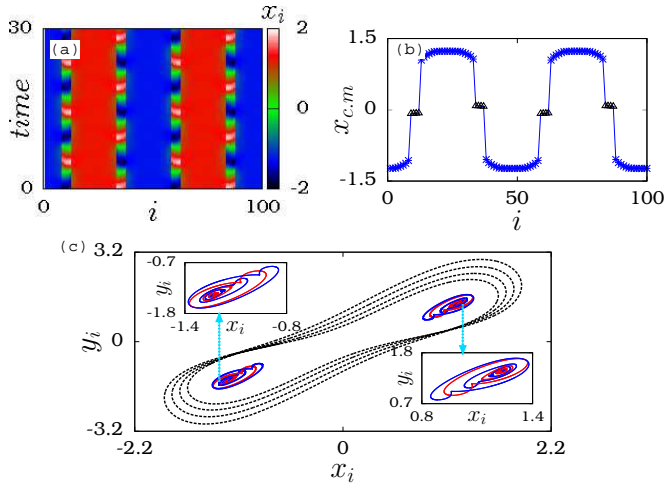}
		\caption{(color online) (a) Space-time plot depicting the oscillatory cluster ($OC$) state for $\alpha=0.5$, $\epsilon=0.95$ and $q=1.0$.  Center of mass of each oscillator is shown in  (b) Here, the center of mass is calculated by averaging over one period of each oscillator. Center of mass of homogeneous oscillations are denoted by unfilled   triangles and star   represents the oscillators in the inhomogeneous group. (c) Phase portrait of representative oscillators from homogeneous and inhomogeneous group of oscillators. Enlarged images of the inhomogeneous oscillators are shown in the insets.}
		\label{sand}
	\end{figure}
	
	\begin{figure}[htb!] 
		\centering
		\hspace{0.1cm}
		\includegraphics[width=8.50cm]{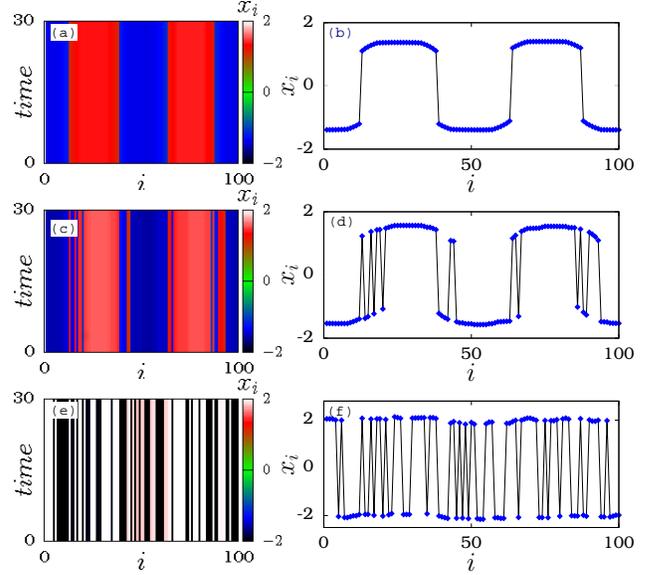}
		\caption{(color online) Space-time and snapshot images for $\alpha=0.5$ and q=1.  (a) two cluster oscillation death  ($2C-OD$) for  $\epsilon=1.05$, (b) two cluster chimera death ($2C-CD$) for $\epsilon=1.18$  and  (c) multi cluster chimera death ($MCD$) for $\epsilon=2.0$.}
		\label{od}
	\end{figure}
	\begin{figure}[htb!] 
		\centering
		\hspace{0.1cm}
		\includegraphics[width=8.50cm]{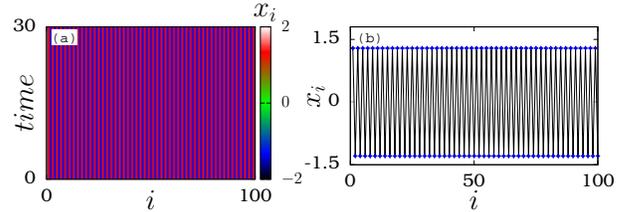}
		\caption{(color online) Space-time and snapshot images of incoherent oscillation death $(IOD)$ state for $\alpha=0.5$, $q=1$   and  $\epsilon=2.0$, }
		\label{iod_ss}
	\end{figure}
	\par Splitting of the desynchronized oscillators into  homogeneous and inhomogeneous ones  facilitating the onset of oscillatory clusters (clusters of different 
	oscillatory states) was observed for further increase in the strength of the nonlocal coupling $\epsilon$.  The spatio-temporal plot of an oscillatory cluster ($OC$)  is 
	depicted in Fig.  \ref{sand}(a).  In order to differentiate the domains of homogeneous and inhomogeneous states we have calculated the  center of mass using 
	the formula $ x_{c.m.}= \int_{0}^{T} \frac{x_i(t) dt}{T}$, where $T=\frac{2\pi}{\omega}$ is the period of oscillation. The center of mass of each oscillator  is depicted 
	in Fig. \ref{sand}(b), where the stars  represent inhomogeneous oscillations and the unfilled triangles denote the homogeneous oscillations.  From the 
	center of mass analysis, it is clear that the homogeneous oscillations always oscillate about the origin characterized by null value of the center of mass whereas the 
	inhomogeneous states  take the center of mass away from the origin  characterized by nonzero values of the center of mass. We  have depicted the phase portraits 
	of homogeneous and inhomogeneous oscillators in the oscillatory  clusters in Fig. \ref{sand}(c) and the enlarged images of inhomogeneous groups are shown in the inset.  
	It is evident that the homogeneous group  oscillates with larger amplitudes (represented by dashed lines in Fig. \ref{sand}(c)) than the inhomogeneous group (solid lines). Further the inhomogeneous group contains more number of oscillators than the homogeneous group.   Increasing   $\epsilon$ further, the number of oscillators in the homogeneous group decreases thereby leading to an increase in the number of oscillators in the inhomogeneous oscillatory group and finally ending up with only a  stable inhomogeneous steady state in the strong coupling limit. The space-time and snapshot images of two coherent cluster steady states ($2C-OD$) are depicted in Figs. \ref{od}(a)-(b) for $\epsilon= 1.05$. The oscillators in the coherent edges moves randomly between the upper and the lower branches of the $IHSS$ for further larger $\epsilon$ resulting in  two cluster chimera death ($2C-CD$) state as shown in  Figs. \ref{od}(c)-(d).   While increasing the strength of the  coupling beyond  $\epsilon=1.2$,  stable multi-chimera death state ($MCD$) emerges from $2C-CD$ (see Figs. \ref{od}(e)-(f)).
	
	\par Stable incoherent oscillation death $(IOD)$ state was also  found to coexist in the region of stable chimera death state upon distributing the initial state of the oscillators at nearly incoherent oscillation death state.  The space-time and snapshot images of $IOD$ state is shown in Figs. \ref{iod_ss}(a)-(b) for $\epsilon=2.0$. In this state, the oscillators occupy the upper and the lower branches of inhomogeneous steady state alternately as depicted in Fig. \ref{iod_ss}(b).  Thus it is evident that the trade-off between the repulsive and attractive nonlocal couplings facilitates 
	the onset of a  rich variety of collective dynamics in  a network of vdP oscillators.  In particular, the competing effects between both the couplings lead to a new transition route to the chimera  death state, namely  oscillatory cluster state.  In the earlier reports, the chimera death was observed through a transition from amplitude chimera to chimera death via in-phase synchronized state for lower range of coupling strengths and a direct transition from amplitude chimera to chimera death was reported at larger coupling strengths. The transition routes were reported with respect to coupling range($r$) \cite{cd1}. The amplitude chimera dynamics  consists  of coherent homogeneous and incoherent inhomogeneous oscillations. The inhomogeneous oscillations in the incoherent state occupy upper and lower branches alternately with different amplitudes. Instead, here the chimera death is observed through oscillatory cluster state  with respect to the strength of repulsive coupling $q$.  Here, both homogeneous and inhomogeneous oscillations are in the form of clusters. Moreover, the amplitude chimera dynamics was reported  as transient in the previous reports  whereas in this work the oscillatory cluster dynamics is found to be stable in all the range of coupling radius \cite{cd3}. The emergence of oscillatory cluster as a function of the  coupling range  $r$ and the strength of nonlocal repulsive coupling $\epsilon$ will be discussed in the following section. 
	
	\subsection{Oscillatory cluster state with respect to $q$ and $r$}
	
\begin{figure}[htb!] 
		\centering
		\includegraphics[width=7.00cm]{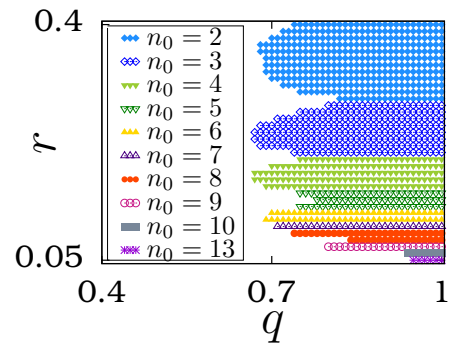}
		\caption{(color online) (a) The number of clusters (coherent domains) of oscillatory cluster state in ($q,r$) plane for coupling strength $\epsilon=0.97$ and $\alpha=0.5$. }
		\label{error}
	\end{figure}

	To understand the robustness of $OC$ state with respect to the coupling range $r$, we have depicted the number of coherent clusters (homogeneous oscillatory states) in the $(q,r)$ plane   in Fig. \ref{error} for a fixed value of $\epsilon=0.97$.   It is evident from the figure that the oscillatory cluster states emerge only above a critical coupling strength of the repulsive interaction.  Different symbols attribute to distinct number of oscillatory clusters with respect to the nonlocal coupling radius $r$. It is also clear that the number of coherent clusters decreases while their spread increases as a function of the coupling range $r$.
	
\begin{figure}[htb!] 
		\centering
		\hspace{-1.1cm}
		\includegraphics[width=7.0cm]{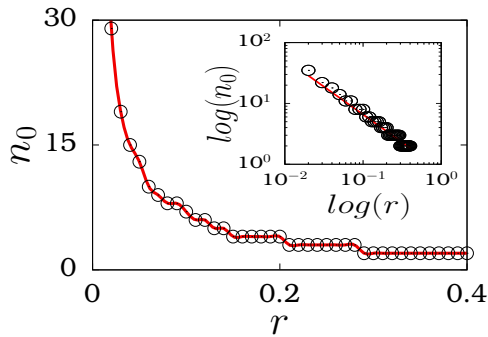}
		\caption{(color online) Exponentially decreasing   number of clusters with respect to coupling range  for $N=500$.  The corresponding power law fit is shown in the inset
			with logarithmic scale.  Unfilled circles represent the numerical data and the corresponding best fit is shown by red solid line. }
		\label{fit}
	\end{figure}

	There is an exponential decrease in the number of coherent clusters in the oscillatory cluster state as depicted in Fig. \ref{fit}  for $N=500$. The inset in Fig. \ref{fit} represents log-log plot illustrating that the number of coherent domains obeying the power law $n_0=ar^b$ with respect to coupling range $r$.  We have also found best fit for $a= 0.802725$ and $b=-0.913692$ which is represented by the red solid line while the corresponding numerical  data is represented by  the unfilled circles in Fig. \ref{fit}.    
	\section{global dynamical behaviour with respect to the strength of the repulsive coupling $q$ }
	\begin{figure*}[htb!] 
		\centering
		\hspace{-2.500cm}
		\includegraphics[width=15.60cm]{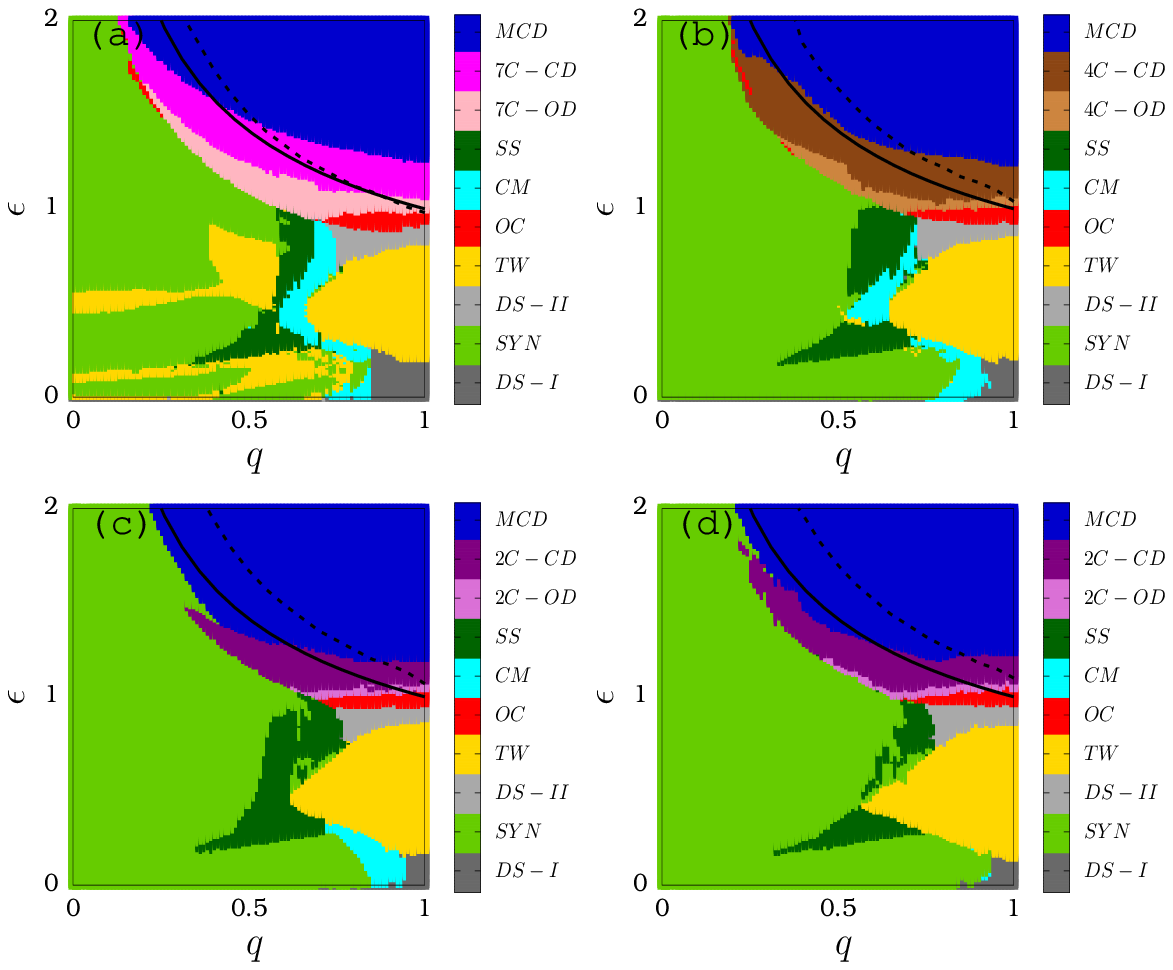}
		\caption{(color online) Two parameter diagrams in ($q, \epsilon$) space for $\alpha=0.5$ and for the coupling range values (a) $r=0.1$,  (b) $r=0.2$, (c) $r=0.3$ and (d) $r=0.4$. Different colors delineat distinct dynamical regimes.   $CM, SS, SYN$ and $DS$ $(I \&  II)$ represent chimera, solitary,  synchronized  and desynchronized states($I$  \&  $II$). $TW$ and $OC$ denote travelling wave and oscillatory cluster states. $2C-OD, 4C-OD$ and $7C-OD$ describe two, four  and seven cluster oscillation deat0h states.  Analogously  $2C-CD, 4C-CD$ and $7C-CD$ are the two, four and seven cluster chimera death states.  $MCD$ describes multi-chimera death state. The dotted line  and solid line represent the Floquet stability curve for stable $SYN$ state and linear stability curve  for $IOD$ state, respectively.}  
		\label{qall}
	\end{figure*}

	In order to understand the global dynamical behaviour of the network of coupled vdP oscillators  as a function of the strength of the repulsive coupling $q$ and coupling strength $\epsilon$, we have  plotted  the two parameter plot in $(q, \epsilon)$ space for four distinct coupling radius values $r=0.1$, $r=0.2$, $r=0.3$ and  $r=0.4$  in Fig. \ref{qall}.  Travelling wave ($TW$) and completely synchronized oscillations ($SYN$) emerge alternately for lower values of $q$ and $\epsilon$ (see Fig. \ref{qall}(a))
	for the coupling radius $r=0.1$.   Increasing $\epsilon$ for lower values of $q$ results  only in  completely synchronized state. Solitary state ($SS$), chimera ($CM$)
	and travelling wave $(TW)$ state  emerges as a function of $q$ for $\epsilon<1$ and $OD$ state emerges for larger values of $\epsilon$ ($\epsilon>1$). For the  strength of the repulsive coupling $q = 1$,  desynchronized state ($DS-I$) is observed at  very low values of $\epsilon$ and travelling waves for larger values of $\epsilon$. Further, transition to desynchronized state ($DS-II$), oscillatory cluster state ($OC$), seven cluster oscillation  death ($7C-OD$), seven cluster chimera death ($7C-CD$) and multi-chimera death states ($MCD$) are observed in Fig. \ref{qall}(a) as a function of the strength of the nonlocal coupling $\epsilon$ for  $r=0.1$.

	Now, we increase the coupling range $r$ from $0.1$ to $0.2$  and  the corresponding two parameter phase diagram is depicted in Fig. \ref{qall}(b).   Emergence of travelling wave  and synchronized state alternately for $r=0.1$ is suppressed  for $r=0.2$ thereby rendering the oscillators to evolve in complete synchrony for lower values of $q$ and $\epsilon$ (see Fig. \ref{qall}(b)). One  can observe  interesting collective dynamics  only above a moderate value of q $(q >0.5)$.   In this  coupling range,   similar dynamical behaviours are observed as in Fig. \ref{qall}(a)  for  $r=0.1$ except for the four cluster oscillation death ($4C-OD$) state and four cluster  chimera death ($4C-CD$) state.   Figures \ref{qall} (c) and \ref{qall}(d) depict the dynamical behavior for the coupling ranges $r=0.3$ and $r=0.4$, respectively.  Initially for lower  values of $q$  synchronized state is observed for all values of $\epsilon$.   At moderate values of $q$ $ (q\approx0.5)$, the  coupled system exhibits synchronized behavior for lower 
	values of $\epsilon$.   There is  an excursion of  some isolated oscillators away from the synchronized group leading to solitary state for further increase in $\epsilon$. The  coupled oscillators exhibit completely synchronized state upon increasing $\epsilon$ further.  In this region,  swing like behaviour of synchronized state is observed.   
	At strong coupling limits ($\epsilon>1$ and $q \approx 0.5$),  the coupled oscillators exhibit a steady state behavior.  There is a direct transition from completely synchronized oscillatory state  to steady state due to  strong trade-off between the attractive and repulsive couplings.  By increasing $\epsilon$, travelling wave ($TW$) state  is found to be stable upto $\epsilon \approx 0.8$.  Beyond this the oscillators become unstable and the coupled system evolves  desynchronously. Further increase in $\epsilon$ results in stable homogeneous and inhomogeneous states thereby facilitating oscillatory cluster state ($OC$).  Finally the coupled system attains steady state where all the oscillators reach inhomogeneous state.  For the coupling range $r=0.3$, the two cluster ($2C-OD$) oscillation death state emerges and becomes two cluster chimera death state as a function of the nonlocal coupling strength $\epsilon$.   
	In $2C-CD$ state the oscillators in two coherent edges  jump between upper and lower inhomogeneous branches facilitating two coherent states and two incoherent states.   
	At strong coupling limits the two cluster chimera death ($2C-CD$) state becomes a multi-chimera death state ($MCD$), which is characterized by the emergence of multiple 
	coherent and incoherent branches of steady states.  Similar  dynamical behaviors  are observed in Fig. \ref{qall}(d) for $r=0.4$  except for the emergence of chimera state, 
	oscillatory cluster and chimera death states.  It is also to be noted that the transition from desynchronized state to chimera death state is always observed through the oscillatory cluster ($OC$) states for all values of the coupling radius $r$ in contrast to the reports in the literature, where chimera death is observed due to nonisochronicity parameter.  In general, complex collective dynamics and their dynamical transitions are observed only for larger values of the repulsive coupling strength $q>0.5$ elucidating its importance in  inducing complex collective dynamics compared with the counter-active attractive coupling.

	\par In addition,  incoherent oscillation death state $(IOD)$ coexists with chimera death  state in certain regions of parameter space. The analytical boundary of $IOD$ state is deduced  from the following linear stability analysis. By distributing the initial state of the oscillators  nearly at the inhomogeneous steady state,  the system exhibits incoherent oscillation death state.  At this $IOD$ state, the system has the following fixed points, $(x_{i-P},y_{i-P})= (x_0,y_0)$, .., $(x_{i-2},  y_{i-2})=(x_0,y_0)$, $(x_{i-1},y_{i-1})=(-x_0,-y_0)$, $(x_i,y_i)=(x_0,y_0) $, $(x_{i+1},y_{i+1})=(-x_0,-y_0)$, $(x_{i+2},y_{i+2})=(x_0,y_0)$..,  $(x_{i+P},y_{i+P})=(x_0,y_0)$.
	By substituting the above mentioned fixed points in Eq. (\ref{model}),  the system equation can be reduced as
	\begin{eqnarray}
	\centering
	{y}_{0} - \beta x_0 &=& 0, \nonumber \\ 
	\alpha (1-{ x}^2_{0}) { y}_{0} -{ x}_{0} +q\beta{ y}_{0}&=&0,
	\label{fix1}
	\end{eqnarray}
	
	where, $\beta = \epsilon $ for even number of nearest neighbours and $\beta = (\frac{P+1}{P})q\epsilon $ for odd number of nearest neighbours.  The explicit fixed point  solutions for Eq. (\ref{fix1}), which can be deduced as 
	\begin{eqnarray}
	x_0 &=&  \pm \sqrt{ \frac{-1 + \alpha \beta+q \beta^2}{\alpha\beta}}, \nonumber \\
	y_0 &=&  {\beta x_0}.
	\label{fp}
	\end{eqnarray}
	The stability condition can be derived by substituting the fixed points ($ x_0, y_0$) in the $2N\times2N$ Jacobian matrix of the system  (\ref{model}) can be expressed as
	
	\begin{align}
	\setcounter{MaxMatrixCols}{20}
	\mathbf{J} & =\begin{pmatrix}
	& a_{11} & a_{12} & a_{13} &  a_{14} &  \cdots& a_{1N} & \\
	& a_{21} & a_{22} & a_{23} &  a_{24} &  \cdots& a_{2N} &\\
	& a_{31} & a_{32} & a_{33} &  a_{34} &  \cdots& a_{3N} &\\
	& a_{41} & a_{42} & a_{43} &  a_{44} &  \cdots& a_{4N} &\\
	&\vdots & \vdots & \vdots & \vdots   & \ddots  & \vdots & \\
	& a_{d1} & a_{d2} & a_{d3} &  a_{d4} &  \cdots& a_{dN} & 	
	\end{pmatrix}
	\end{align} 
	with $a_{ii}=-\epsilon$, $a_{i(i+1)}=1$ and  $a_{i(i+j+1)}=a_{i(i-j)}=\frac{\epsilon}{2P}$ for $i=1,3..,{(N-1)}$ and ${j=1,3..,{(2P-1)}}$.  Analogously, $a_{i(i-1)}=\gamma_1$, $a_{ii}=\gamma_2$  and  $a_{i(i+j)}=a_{i(i-j)}=\frac{-q \epsilon}{2P}$ for $i=2,4..,N$ and $j=2,4.., 2P$ where $\gamma_1= 1-2\alpha\beta-2q\beta^2$ and $\gamma_2=\frac{1}{\beta}-q\beta+q\epsilon$. Here, $a_{i(N+j)}=a_{ij}$ and $a_{i(1-j)}=a_{i(N-j+1)}$ for $j=1,2..,P$.  From the eigen value analysis conditions for  stable $IOD$ region can be  obtained \cite{clust_od}. In the stable IOD region the following condition must be satisfied, 
		\begin{figure*}[htb!] 
			\centering
			\hspace{-1.1cm}
			\includegraphics[width=15.00cm]{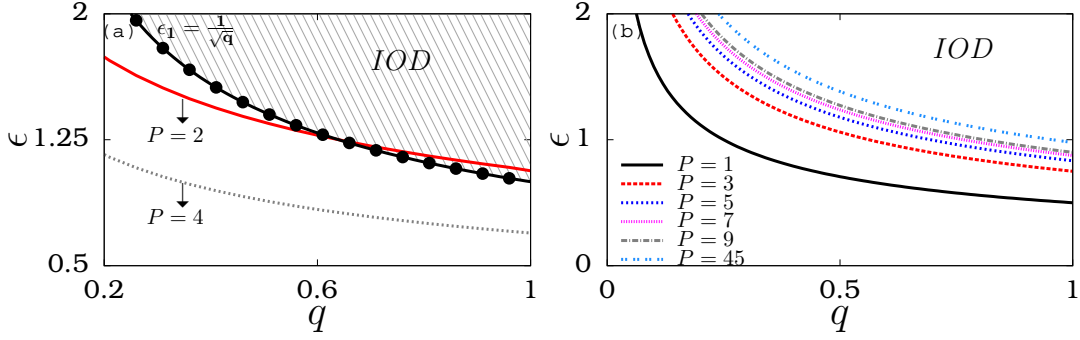}
			\caption{(color online) Boundaries  of incoherent  oscillation death in (q, $\epsilon$) space for (a) even $P$  and (b) odd $P$. Stability curves with respect different  nearest neighbours are denoted by distinct line types.   The shaded region in Figure (a) represents the stable region of $IOD$ state for even number of $P$.  }
			\label{iod}
		\end{figure*}
	\begin{eqnarray}
	Tr(J) = \lambda_1+\lambda_2+\lambda_3+...\lambda_{2N} < 0.
	\end{eqnarray}
	For odd and even number of nearest neighbours, the trace of  $2N \times 2N$ Jacobian matrix of the considered system can be expressed as 
	\begin{eqnarray}
	Tr(J) = N(\alpha c_1-\epsilon \hat{q}) <0. 
	\label{tr} 
	\end{eqnarray}
	The following additional condition has also to be satisfied for fixed points to be stable, 
	\begin{eqnarray}
	det(J) = \lambda_1 \, \lambda_2 \,\lambda_3 \,...\,\lambda_{2N}>0.
	\end{eqnarray}
	
	The stability  conditions are found by equating the eigen values of the determinant to zero. In the stable $IOD$ region real parts of all the eigen values acquire negative values.	Mainly,  the following  eigen values play crucial role in determining the stability,
	\begin{eqnarray}
	\lambda_{1}&=& \frac{1}{4}(2\alpha c_1\pm \sqrt{(-2\alpha c_1)^2-16c2}) \\
	\lambda_{2}&=& \frac{1}{4P}(2P\alpha c_1-6\epsilon\hat{q} \nonumber \\
	&& \pm  \sqrt{(-2P\alpha c_1+6\epsilon\hat{q})^2-16(P^2c_2-3P\alpha\epsilon c_1-9q\epsilon^2)}) \nonumber \\
	\label{lam2}
	\end{eqnarray}
	where, $c_1=(1-x^2)$ and $c_2 = 1+2xy\alpha$ and $\hat{q}=1-q$.  The stable region emerges at,
	\begin{center}
		$\epsilon_1=\frac{1}{\sqrt q}$:  for even $P$ \\ $\epsilon_2=(\frac{P}{P+1})\frac{1}{\sqrt q}$: for odd $P$. 
	\end{center}
	which are obtained from $\lambda_{1}$. The stability condition deduced from  $\lambda_1$ can be denoted as $\epsilon_1$ and $\epsilon_2$. Similary the stability condition deduced from  $\lambda_2$ can be represented by $\epsilon_3=(\frac{\epsilon k_2-2Pk_1}{\epsilon}+\sqrt{4P(-P-3\alpha\epsilon+4k_1)-36 q \epsilon^2 +(\frac{2Pk_1}{\epsilon}-k_2)^2})$,  where $k_1 = -1+\alpha \epsilon+q \epsilon^2 $ and $k_2 = 2P\alpha-6 \epsilon+6q\epsilon $. The stable $IOD$ region is enclosed by $\epsilon_1>\frac{1}{\sqrt q}$ for $q<0.65$  and  $\epsilon_3>\frac{-4\alpha+\sqrt{2}\sqrt{25 q +8 \alpha ^2}}{5\sqrt{q}}$ for $q>0.65$ $(N=6, 8, 10,12..,100)$.  For $P=2,$  the stable region satisfies both the above conditions.   The corresponding stability curves  enclosing the stability regions of $IOD$ state for even number of nearest neighbours  is depicted in Fig. \ref{iod}(a). The solid line and dashed line  in Fig. \ref{iod}(a) represent  $\epsilon_3$  at $P=2$ and $P=4$, respectively.  Comparing both the curves  it is clear that for large  $P$, $\epsilon_3$ curve emerges below the stable region.  Further for all other even $P$, the stable region is enclosed only by $\epsilon_1>\frac{1}{\sqrt q}$. The stability curves $\epsilon_2$ enclosing the stable IOD state for odd $P$'s are depicted in Fig. \ref{iod}(b). The distinct lines correspond to different values of nearest neighbours elucidating that  the stable region depends on the number of  nearest neighbours.  It is to be noted that the stable region decreases with increase in number of  nearest neighbours.  
	\begin{figure}[htb!] 
		\centering
		\hspace{-1.1cm}
		\includegraphics[width=7.50cm]{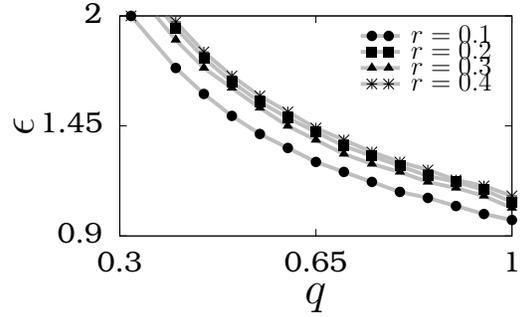}
		\caption{(color online) Boundaries of stable completely synchronized state  for four  distinct  coupling ranges in (q, $\epsilon$) space.  Different line points   such as circle, rectangle, triangle and star points with line represent coupling ranges $r=0.1, r=0.2, r=0.3$ and $r=0.4$, respectively. }
		\label{flo}
	\end{figure}
	\par The analytical boundary of $IOD$ state  is depicted by solid line in all the two parameter diagrams in Fig. \ref{qall}. The area above the solid line represents the stable $IOD$ region satisfying the stability condition  $\epsilon \ge \frac{1}{\sqrt{q}}$ for $q \le 1$, which is valid  for all even number of nearest neighbours.   It is to be noted that the above stability condition is the same as that of the two coupled network as it repeats for $N$- coupled network  with even number of nearest neighbours.
	\par  Finally, the area under the dashed line in all the  two phase diagrams denote stable regions of sychronized state which is confirmed using Floquet exponents 
	from  Eq. (\ref{floq}). The stable synchronized oscillations coexists with $SS$, $TW$, $CM$, $DS-I$, $DS-II$ and $OC$ states.  The synchronized state  also  coexists with 
	coherent oscillation death and chimera death states in certain ranges of parameters. The stability curves of synchronized states are shown separately in Fig. \ref{flo},  where the  four different line points represent the four distinct coupling radius $r=0.1$, $0.2$, $0.3$ and $0.4$.  The stability of synchronized state was also verified through master stability function (MSF) \cite{msf}-\cite{gopal_msf}.

	\section{Conclusions}
	In summary, we have investigated the emergence of chimera death states in a network of nonlocally coupled van der Pol oscillators due to the combined effect of  both the  attractive and repulsive couplings.  We found that the repulsive coupling plays a crucial role in inducing various collective dynamics.  In particular, we have shown that the chimera death state emerges through a novel  transition route, namely,   oscillatory cluster state $(OC)$  due to strong trade-off between the attractive and repulsive couplings.  It is also shown that the network of coupled   vdP oscillators  takes the same transition route in all  the coupling ranges. Strong competing interaction between the couplings renders the oscillators either in homogeneous or inhomogeneous  oscillatory states. We have also elucidated that an increase in the coupling radius  leads to a decrease  in the number of clusters (coherent states) and increases  the cluster size (coherent regions). Finally we have observed that increase in  coupling strength leads to structural changes in coherent death state and to the transition from  coherent oscillation death to multi-chimera death via distinct cluster chimera death state. Furthermore the coexistence of incoherent death  state is also found to coexist with the chimera death state and the corresponding analytical stability was deduced.  Analytical stability curves for the synchronized and  solitary states were also deduced.
	
	\section*{Acknowledgement}
	KS wishes to thank SASTRA University for research fund and extending infrastructure support to carry out this work. DVS is supported by the CSIR EMR Grant No. 03(1400)/17/EMR-II. The work of VKC forms part of a research project sponsored by INSA Young Scientist Project under Grant No. SP/YSP/96/2014.  
	The work of ML is supported by a DST-SERB Distinguished Fellowship program.


\begin{thebibliography}{99}	
		\bibitem{int} A. Pikovsky, M. G. Rosenblum, and J. Kurths, Synchroniza-
		tion, A Universal Concept in Nonlinear Sciences (Cambridge
		University Press, Cambridge, 2001).
		\bibitem{od} W. Zou, D. V. Senthilkumar, R. Nagao, István Z. Kiss, Y. Tang, A. Koseska, J. Duan and J. Kurths, Nat Commun. 6: 7709 (2015)
		\bibitem{od_r1} D. G. Aronson, G. B. Ermentrout, and N. Koppel, Physica D {\bf 41}, 403 (1990).
		\bibitem{od_r2} P. C. Matthews and S. H. Strogatz, Phys. Rev. Lett. {\bf 65}, 1701 (1990); P. C. Matthews, R. E. Mirollo and S. H. Strogatz, Physica D {\bf 52 }, 293 (1991).
		\bibitem{od_r5} A. Sharma and M. D. Shrimali, Phys. Rev. E. {\bf 85}, 057204
			(2012).
		\bibitem{od_r4} A. Koseska, E. I. Volkov, A. Zaikin, and J. Kurths, Phys.
		Rev. E. {\bf 75}, 031916 (2007).
		\bibitem{add}D. V. Reddy, A. Sen, and G. L. Johnston, Phys. Rev. Lett. {\bf80}, 5109–5112 (1998).
		\bibitem{iprl1968} I. Prigogine, R. Lefever,   J. Chem. Physics {\bf48}, 1695?1700 (1968); Koseska, A., Volkov, E. \& Kurths, J.   \textit{Phys. Rep.} \textbf{531}, 173-199 (2013).
		\bibitem{od_sen1} Wei Zou, D. V. Senthilkumar, Yang Tang, and J\"urgen Kurths,  Phys. Rev. E. {\bf 86}, 036210 (2012).
		\bibitem{od_sen2}Wei Zou, D. V. Senthilkumar,  Jinqiao Duan, and J\"urgen Kurths,  Phys. Rev. E. {\bf 90}, 032906 (2014).
		\bibitem{od_rr4} K. Konishi, Phys. Rev. E {\bf 68}, 067202 (2003).
		\bibitem{od_rr5} R. Karnatak, R. Ramaswamy and A. Prasad, Phys. Rev. E {\bf 76}, 035201(R) (2007).
		\bibitem{od_rr6} V. Resmi, G. Ambika, R. E. Amritkar Phys. Rev. E {\bf 84}, 046212 (2011).
		\bibitem{od_rr7} C. R. Hens, O. I. Olusola, P. Pal and S. K. Dana, Phys. Rev. E {\bf 88}, 034902 (2013).
		\bibitem{ull}E. Ullner, A. Zaikin, E. I. Volkov, and J. Garcia-Ojalvo,
		Phys. Rev. Lett. {\bf 99}, 148103 (2007).
		\bibitem{od_app1} M. F. Crowley and I. R. Epstein, J. Phys. Chem. {\bf  93}, 2496
		(1989).
		\bibitem{od_app2}M. Toiya, V. K. Vanag, and I. R. Epstein, Angew. Chem.,
		Int. Ed. {\bf 47}, 7753 (2008).
		\bibitem{od_app3} K. P. Zeyer, M. Mangold, and E. D. Gilles, J. Phys. Chem. A
		{\bf 105}, 7216 (2001).
		\bibitem{od_app4} M. Heinrich, T. Dahms, V. Flunkert, S. W. Teitsworth, and
		E. Sch\"oll, New J. Phys. {\bf 12}, 113030 (2010).
		\bibitem{clust_od} A. Mishra, S. Saha, P. K. Roy, T. Kapitaniak, and S. K. Dana, Chaos {\bf 27}, 023110 (2017).
		\bibitem{chi1}Y. Kuramoto and D. Battogtokh, Nonlinear Phenom.
		Complex Syst. 5, 380 (2002).
		\bibitem{chi2} D. M. Abrams and S. H. Strogatz, Phys. Rev. Lett. 93,
		174102 (2004).
		\bibitem{chi3} D. M. Abrams and S. H. Strogatz, Int. J. Bif. Choas
		16(1), 21 (2006).
		\bibitem{chi4}M. J. Panaggio and D. M. Abrams, Nonlinearity 28, R67
		(2015).
		\bibitem{t2} I. Omelchenko, A. Provata, J. Hizanidis, E. Sch\"öll, and
		P. H\"ovel, Phys. Rev. E {\bf 91}, 022917 (2015).
		\bibitem{t3}  T. Banerjee,  P. S. Dutta,  A. Zakharova,  and E. Sch\"oll,
		Phys. Rev. E {\bf 94}, 032206 (2016).
		\bibitem{t4} I. Omelchenko, Y. Maistrenko, P. H\"ovel, and E. Sch\"oll,
		Phys. Rev. Lett. {\bf 106}, 234102 (2011).
		\bibitem{t5}   V. Semenov, A. Feoktistov, T. Vadivasova, E. Sch\"oll, and
		A. Zakharova, Chaos {\bf 25}, 033111 (2015).
		\bibitem{chi_1} Y. Kuramoto and D. Battogtokh, Nonlinear Phenom. Complex Syst. {\bf 5}, 380 (2002).
		\bibitem{chi_2}	D. M. Abrams, and S. H. Strogatz, Phys. Rev. Lett. {\bf 93}, 174102 (2004).
		\bibitem{chi_3} M. J. Panaggio, and D. M. Abrams, Nonlinearity {\bf 28}, 3 (2015).
		\bibitem{gopal1}R Gopal, V. K. Chandrasekar, A Venkatesan, M Lakshmanan, Physical Review E {\bf 89} , 052914 (2014)
		\bibitem{chi_4}K. Kaneko, Physica D  {\bf 41}, 137 (1990); Chaos{\bf 25}, 097608 (2015); A. Yeldesbay, A. Pikovsky, M.Rosenblum,Phys. Rev. Lett. {\bf 112}, 144103 (2014); G. C. Sethia and A. Sen, Phys. Rev. Lett {\bf 112}, 144101 (2014). 
		\bibitem{vkc1}V. K. Chandrasekar, R. Gopal, A. Venkatesan, and M. Lakshmanan, Phys. Rev. E {\bf 90}, 062913 (2014).
		\bibitem{chen1}A. Mishra, C. Hens, M. Bose, P. K. Roy, and S. K. Dana, Phys.Rev.E {\bf  92}, 062920 (2015).
		\bibitem{chi_5} B. K. Bera, D. Ghosh and M. Lakshmanan, Phys. Rev. E {\bf 93}, 012205 (2016).
		\bibitem{chi_6} C. R. Laing, Phys. Rev. E {\bf 92}, 050904(R) (2015).
		\bibitem{chi_7} B. K. Bera and D. Ghosh, Phys. Rev. E {\bf 93}, 052223 (2016).
		\bibitem{chi_8} B. K. Bera, D. Ghosh, and T. Banerjee, Phys. Rev. E {\bf 94}, 012215 (2016).
		\bibitem{a1}  N. C. Rattenborg, C. J. Amlaner, S .L. Lima, Neurosci.
		Biobehav. Rev. 2, 817 (2000).
		\bibitem{a2}  A.  Rothkegel,  K.  Lehnertz,  New  J.  Phys. {\bf 16},  055006 (2014).
		\bibitem{a3}  C.  R.  Laing,  C.  C.  Chow,  Neural  Comput. {\bf 13},  1473 (2001).
		\bibitem{a4} H. Sakaguchi, Phys. Rev. E {\bf 73}, 031907 (2006).
		\bibitem{a5} A. E. Filatova, A. E. Hramov, A. A. Koronovskii, S. Boc-
		caletti, Chaos {\bf 18}, 023133 (2008).
		\bibitem{a6} J.  C.  Gonz ́alez-Avella,  M.  G.  Cosenza,  and  M.  San
		Miguel, Physica A  {\bf 399}, 24 (2014).
		\bibitem{cd1} A. Zakharova, M. Kapeller, and E. Sch\"oll
		Phys. Rev. Lett. {\bf 112}, 154101 (2014). 
		\bibitem{cd2} A. Zakharova, M. Kapeller, and E. Sch\"oll, IOP Publishing, Conference Series {\bf 727},  012018 (2016).
		\bibitem{cd3} L. Tumash , A. Zakharova  , J. Lehnert  , W. Just  and E. Sch\"oll , EPL, {\bf 117}, 20001, (2017). 
		\bibitem{prem1} K. Premalatha, V. K. Chandrasekar, M. Senthilvelan, M. Lakshmanan, Phy. Rev. E {\bf 93} 052213 (2016).
		\bibitem{prem2} K. Premalatha, V. K. Chandrasekar, M. Senthilvelan, and M. Lakshmanan, Phys. Rev. E. {\bf 91}, 052915 (2015).
		\bibitem{qu} Z. Qu, Y. Shiferaw, and J. N. Weiss, Phys. Rev. E
		{\bf{75}}, 011927 (2007).
		\bibitem{neu}T. Yanagita, T. Ichinomiya, and Y. Oyama, Phys. Rev. E
		{\bf{72}}, 056218 (2005).
		\bibitem{gene} Xiao-Jun Tian, Xiao-Peng Zhang, Feng Liu, and Wei Wang, Phys. Rev. E {\bf{80}} 011926 (2009)
		\bibitem{vdp1} J. Guckenheimer, K. Hoffman, and W. Weckesser, Numerical computation of
		canards, Internat. J.Bifur. Chaos Appl. Sci. Engrg., 10 (2000). 
		\bibitem{vdp2}P. F. Rowat and A. I. Selverston, Modeling the gastric mill central pattern generator of the lobster	with a relaxation-oscillator network, J. Neurophysiology, 70 (1993).
	
		\bibitem{daido} H. Daido and K. Nakanishi, Phys. Rev. Lett. {\bf{96}}, 054101, (2006).
		\bibitem{msf} L. M. Pecora and T. L. Carroll, Phys. Rev. Lett. {\bf 80}, 2109, (1998).
		\bibitem{gopal_msf} V. K. Chandrasekar, R. Gopal, D. V. Senthilkumar, and M. Lakshmanan, Phys. Rev. E {\bf 94}, 012208 (2016)
	\end{thebibliography}
\end{document}